\documentclass[twocolumn,preprintnumbers,amsmath,amssymb,showpacs]{revtex4}

\usepackage{graphicx}
\usepackage{dcolumn}
\usepackage{bm}
\usepackage{bbm}
\usepackage{amsmath}

\newcommand{\ket}[1]{|#1\rangle}

\newcommand{\ketbra}[2]{|#1\rangle\!\langle#2|}

\newcommand{\blankonearray}{\begin{array}{c}\phantom{,}\end{array}\phantom{^A{^A}}\!\!\!}

\newcommand{\blankthreearray}{\begin{array}{c}\phantom{,}\\ \phantom{,} \\ \phantom{,}\end{array}\!\!\!\!\!}

\let\oldmarginpar\marginpar
\renewcommand\marginpar[1]{\-\oldmarginpar[\raggedleft\marginparsize #1]%
{\raggedright\marginparsize #1}}

\begin{document}

\setlength{\tabcolsep}{1ex}

\title{The work value of information}

\author{Oscar C. O. Dahlsten*}
\affiliation{Institute for Theoretical Physics, ETH Z\"urich, 8093 Zurich, Switzerland}
\email{dahlsten@itp.phys.ethz.ch}

\author{Renato Renner}%
\affiliation{Institute for Theoretical Physics, ETH Z\"urich, 8093 Zurich, Switzerland}

\author{Elisabeth Rieper}%
\affiliation{Centre for Quantum Technologies, National University of Singapore, 3 Science Drive 2, Singapore 117543, Singapore}

\author{Vlatko Vedral}%
\affiliation{Clarendon Laboratory, University of Oxford, Parks Road Oxford OX1 3PU, United Kingdom\\
Centre for Quantum Technologies, National University of Singapore, 3 Science Drive 2, Singapore 117543, Singapore\\Physics dept., National University of Singapore, 2 Science Drive 3, Singapore 117543, Singapore}



\date{6 May 2009}

\begin{abstract}
 We present quantitative relations between work and information that are valid both
for finite sized and internally correlated systems as well in the thermodynamical limit.
We suggest work extraction should be viewed as a game where the amount of
work an agent can extract depends on how well it can guess the micro-state of the system.
In general it depends both on the agent's knowledge and risk-tolerance, because the agent
can bet on facts that are not certain and thereby risk failure of the work extraction.
We derive strikingly simple expressions for the extractable work in the
extreme cases of effectively zero- and arbitrary risk tolerance respectively, thereby enveloping all cases. Our derivation makes a connection between heat engines and the smooth entropy approach. The latter has recently extended Shannon theory to encompass finite sized and
internally correlated bit strings, and our analysis points the way to an analogous extension
of statistical mechanics. 
\end{abstract}




\maketitle

 \noindent{\bf {\em Introduction}}---The relation between work and information has been the cause of great debate since the beginnings of statistical mechanics. Focal points of  
this debate include Maxwell's demon, Szilard's engine, Landauer's erasure and Bennett's reversible measurements \cite{LeffR02, Szilard29, Landauer61, Bennett82}.    

\begin{figure}[htb]
\centering \includegraphics[width=0.95\linewidth]{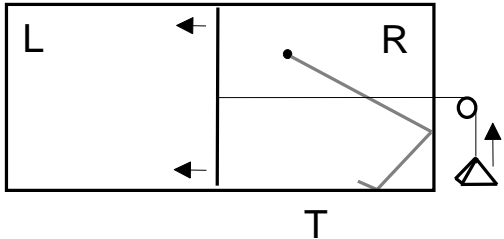}
\caption{Work extraction requires information. In a paradigmatic example, Szilard noted that if one has information that a particle in contact with a thermal reservoir of temperature $T$ is in one given side of a box ($L$ or $R$), one can insert a divider and trap it there~\cite{Szilard29}. The divider is attached to a weight and, as the particle bounces around due to its thermal energy, the divider can be pushed in a predictable direction, lifting the weight. The work output is $kT\ln2$ per such particle, and this is therefore the work value of one bit ($L$ or $R$) in this context.}
\label{fig:box}
\end{figure}

That there should be such a relation can be seen intuitively by noting
that harnessing motion, e.g. wind, for ones benefit requires knowing 
its directionality. In thermodynamical work extraction 
from the pressure of a gas one uses the knowledge that the particles are confined 
and will only push the piston from one known direction. The simplest 
example of such extraction is perhaps Szilard's seminal engine, described 
in Figure~\ref{fig:box}.   

In the context of Szilard's engine, previous efforts to quantify the relation between work and information yielded expressions of the type $W=(n-S)kT\ln2$, where $W$ is the work out, $n$ the number of particles, $S$ the lack of information (entropy) about the positions of individual particles and $nkT\ln2$ the amount of work that would be gained if there were no uncertainty~\cite{Bennett82, Feynman}. Feynman argued this expression {\em defines} entropy $S$~\cite{Feynman}.

In this Letter we revisit this relation in the light of the recently developed smooth entropy approach~\cite{Renner05,RennerW04}.  This approach has enabled the extension of Shannon theory in a simple yet accurate manner so that is also valid for finite sized and correlated bit strings, and it is intriguing to ask whether something analogous can be achieved for statistical mechanics.   
 
We suggest as part of our approach that work extraction should be treated as a game where an agent uses its information to extract work by guessing aspects of the microstate of the system. The agent uses information compressing unitaries as part of this process. The agent also has to choose a trade-off between risk of failure and work extracted if successful. The work value of information therefore depends on the risk tolerance too. To recover a simple theory, we focus on the extreme cases of effectively no- and arbitrary risk tolerance respectively, as these cases envelope all others. We derive the two respective work values of information, and discuss the consequences. We recover the standard result in the appropriate limit, but show the work value(s) can in general be very different from that. The results hold universally for quantum systems and classical systems in the same way information compression bounds do. 

The presentation proceeds as follows. We firstly summarize the smooth-entropy approach. We then describe existing ideas on how to use information to extract work, in particular the idea of using information compression in quantum systems. We go on to define the work extraction game within which to quantify the work value of information. We derive the two statements concerning the work value of information, and then discuss the implications.

\noindent{\bf {\em Smooth entropies}}---Given a probability distribution P with entries $p_i$, or equivalently a density matrix with eigenvalues $\lambda_i$,  there are numerous ways to assign to it a number quantifying the associated ignorance, i.e., entropy. A commonly used function is the Shannon entropy $H=-\sum_i p_i\log(p_i)$ (we shall also denote it by $H_S$). The reader is less likely to be familiar with the {\em max} and {\em min} entropies which shall both be needed here. They are called this because $H_{\min}\leq H \leq H_{\max}$. They are defined by $H_{\max}(P):=\log|\mbox{supp(P)}|$ and $H_{\min}(P):=-\log(\max_i p_i)$ respectively, where $|\mbox{supp(P)}|$ is the size of the support of the distribution, $\max_i p_i$
 is the peak value and the logarithm is to base 2. For the distribution 
$P_{ex}=[0.5 \,\, 0.49998 \,\, 0.00001 \,\, 0.00001 \,\, 0]$
for example, $H_{\max}(P)=\log(4)=2$ and $H_{\min}(P)=-\log(1/2)=1$.

An operational meaning of $H_{\max}$ is that it answers the question of how many bits  (two-level systems) a memory would need in order to store a message from the distribution. $H_{\min}$ on the other hand bounds how many out of the $n$ bits that are unbiased, in the sense that the marginal distribution on them is uniform. The marginal distribution on any number of bits will always have an entry which is at least of size $p_{\max}:=\max_i p_i$. One can accordingly, bearing in mind that the marginal distribution is normalised, not find a marginal distribution that is uniform and has more than $1/p_{\max}$ events. Thus no more than $\log(1/p_{\max})=H_{\min}$ bits can be uniformly distributed. One can moreover say (up to a small term) that $H_{\min}$ bits {\em are} completely unknown, as will later be shown in the proof of Theorem II. 

In practical applications one will normally not care about extremely unlikely events. This motivated the recently suggested  \cite{RennerW04} modified versions of the two entropies. The modified versions are called the {\em smooth} min and max entropy respectively, since they typically do not vary much under small changes in the probability distributions. They are defined in the following manner:\vspace{-0.1cm}
\begin{eqnarray}
H_{\min}^\epsilon (P)&:=&\max_{\overline P } H_{\min}(\overline P),\\
H_{\max}^\epsilon (P)&:=&\min_{\overline P} H_{\max}(\overline P). 
\end{eqnarray}\vspace{-0.1cm}
The maximum/minimum is taken over all $\overline P$ such that the statistical distance $d(P,\overline P)\leq \epsilon$ (the trace distance in the quantum case). The parameter $\epsilon$ can be interpreted as the maximum probability of events one is prepared to ignore in the analysis and is normally taken to be very small, but non-zero.

In line with the definition, with probability $p>1-\epsilon$, a memory of size $H_{\max}^\epsilon$ will be enough to store a string from the distribution correctly. For example, with $p>1-0.00002$ a memory of size $H_{\max}^{0.00002}=\log(2)=1$ bit would suffice for $P_{ex}$ from before.

By the {\em asymptotic equipartition theorem}, both entropies converge to the Shannon entropy for $n$ i.i.d. distributed particles as $n\to \infty$ and $\epsilon \to 0$, see e.g.~\cite{TomamichelCR08}. This is only true for the smooth versions.  

 Readers familiar with the smooth entropy literature can note that what we call $H_{\max}^\epsilon$ here is the smooth Renyi-entropy of order $0$ and not that of order $1/2$, but these only differ by at most $\log(1/\epsilon)$ \cite{RennerW04}. 

\noindent{\bf {\em Szilard's engine and Bennett's development}}---In this work we will consider the work value of information in a quite general work extraction scenario. To understand why we chose this scenario it is instructive to recall certain specific examples existing in the literature. Bennett, in particular, considered $n$ Szilard's engines (like in Figure~\ref{fig:box}) together extracting work from a heat bath~\cite{Bennett82}. The experimenter's knowledge is encoded in the probability distribution on particle positions $\{L,R\}^n$. Each box has a work value $c=kT\ln2$ associated with knowing $L$ or $R$ perfectly. Thus if all boxes are either fully known or completely unknown, $W=(n-n_u)kT\ln2$, where $n_u$ is the number of completely unknown boxes. Bennett notes, crucially, that correlations can be exploited too, even if the marginal distributions on the bits are uniformly random. The experimenter can implement a reversible interaction between the boxes to compress the total randomness/information (which is constrained by the correlations) onto individual bits, so that the others can then be used for work extraction. As a simple example, let $n=2$, $p(LL)=p(RR)=1/2$. Then performing the reversible so-called  controlled-not interaction~\cite{CNOT} would yield $p(LL)=p(RL)=1/2$ so that the second box could be used to extract $c$ work~\cite{Feynman}.  
 
The considerations, like many other information theoretical arguments, apply also to quantum systems \cite{Zurek03}. In fact, the nano and quantum regimes should be the focus for implementations of these ideas, due to the probably unavoidable presence of friction in macroscopic systems. We then replace the $n$ bits with $n$ qubits, and the distribution on $\{L,R\}^n$ with a density matrix $\rho$ representing the state of the $n$ qubits \cite{Zurek03}.  The reversible interactions compressing the information are then unitaryas discussed e.g. in \cite{OppenheimHHH02} and in a closely related setting in \cite{SchulmanV99, FernandezLMR04}. We may assume $\rho$ is diagonal, i.e. a classical distribution, because the experimenter would otherwise apply a unitary to diagonalise it to minimise the uncertainty in the basis in question. 

\noindent{\bf {\em Our work extraction scenario}}---We now construct a work extraction game from the above examples, wherein an agent tries to extract as much work as possible given its information. We say the agent succeeds in extracting work if and only if it lifts a weight (or does work against an analogous counter-force) to a predetermined level. We assume, as is common, that the relative thermal fluctuations in the piston are negligible because all particles are working on the same piston---see \cite{Berger90} for more discussion. The game is defined so that the agent:
\begin{itemize}
\vspace{-0.1cm}
\item Is given $n$ bits/qubits of work value $c$ and a distribution on $\{L,R\}^n  /\{\ket{L},\ket{R}\}^n $.
\vspace{-0.1cm}
\item Presets a unitary to be applied to the particles. 
\vspace{-0.1cm}
\item Presets which of the boxes to use for work extraction after the unitary.
\vspace{-0.1cm}
\item Presets the weight to be lifted.
\vspace{-0.1cm}
\item Then interferes no more in the extraction.
\end{itemize}
\begin{figure}[t]
	\begin{center}
\centering \includegraphics[width=0.8\linewidth]{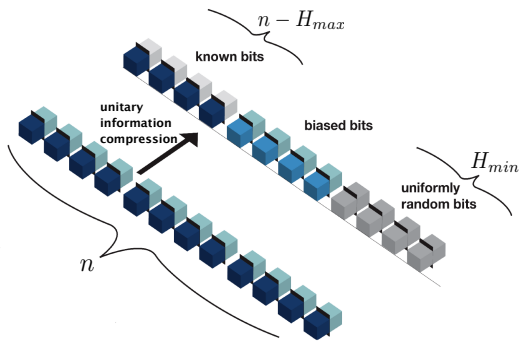}
\caption{Unitary interaction compressing the information such that some bits are made fully known, some are biased and some uniformly random.  Darker colour indicates higher probability density. Each bit represents a box such as in Figure~\ref{fig:box}. The agent can use this process to remove or minimise fluctuations in the work output since known bits will not yield any fluctuations. Note that it is not always obvious whether a box should be coupled to the weight or not. If the box is only biased but not certain, then a risk-averse agent would not use it, but another agent may.}

\label{fig:permutation}
\end{center}
\end{figure}
\noindent{\bf {\em The work value of information}}---The above is a well-defined, physically concrete, and quite general scenario within which to quantify the work value of information. 
  The agent has several choices to make. It is natural to assume the experimenter chooses the most information compressing unitary.   Using uncertain bits and choosing a heavier weight both increase the possible yield and the probability of failure. To simplify the situation we focus on the extreme cases of an agent accepting effectively no risk of failure, and another accepting effectively arbitrary risk of failure. The following Theorems give the work value of information in the two respective cases and hold for single realizations.
 
{\bf Theorem I}: {\em  Except with $p<\epsilon$, the agent can be certain to extract 
\vspace{-0.2cm}
\begin{equation*}
\vspace{-0.1cm}
W=\left(n-H_{\max}^\epsilon \right)c
\vspace{-0.1cm}
\end{equation*}
 work, and no more}. 

{\bf Theorem II}: {\em Except with $p<2\epsilon$, for an agent willing to risk failing to extract work,} 
\vspace{-0.2cm}
\begin{equation*}
\vspace{-0.2cm}
W\! \le \! \left( n\! -\! H_{\min}^\epsilon\! + \! \log \left(\frac{1}{\epsilon}\right)  \right)c.
\end{equation*}

We proceed to outline the proof of the Theorems, omitting some tedious but
straightforward calculations for clarity.

Theorem I follows from the following argument. By a standard smooth entropy result no fewer than $H_{\max}^\epsilon$ bits can be uncertain (except with $p<\epsilon$) so an agent unwilling to use any uncertain bits cannot extract more than  $(n-H_{\max}^\epsilon)c$ work. That the agent can in fact extract that amount of work with certainty follows from noting that the agent can apply the unitary which takes the initial distribution to a state $[p_1, ...p_k,0...0]$ where $k$ is the size of the support of the post-smoothing distribution. Then only $H_{\max}^\epsilon=\log k$ of the bits are uncertain, and the agent can use the remaining ones to extract work. That concludes the proof of Theorem I.

We now proceed to prove Theorem II. We prove, crucially, that for the agent to guess all bits it is using successfully with $p > \epsilon$, it has to desist from using at least $H_{\min} + \log \epsilon$ bits. To see this, note that $\overline{p}_{\max} \geq 2^{(n-\overline{n})}p_{\max}$, where $\overline{p}_{\max}$ is the peak probability of the marginal distribution on the subset. Since $p_{\max} > \epsilon$, $\overline{n} \leq n - \left(H_{\min} + \log \epsilon\right)$. Thus at least $H_{\min} + \log \epsilon$ bits have to be traced out to get $p > \epsilon$ chance of guessing the remaining bits correctly, in which case $W\leq \! \left( n\! -\! H_{\min}\! + \! \log \left(\frac{1}{\epsilon}\right)  \right)c$. It can moreover be shown that the agent cannot exceed this by using an even larger set of bits, nor by varying the counterweight (under the restriction $p>\epsilon$). Finally, to recover the smooth version, we go through the same derivation for an $\epsilon$-close distribution, yielding Theorem II, which accordingly only holds with $p<2\epsilon$.

\noindent {\bf {\em Discussion}}---Four illustrative examples of distributions are discussed in Table~1.

%

\begin{table}
\begin{tabular}{|l|l|l|}
\hline
{\bf Distribution} $\blankonearray$&  {\bf Min work (I)}  & {\bf Max work (II)}\\
\hline 
$ \left[\begin{array}{c}
p(L) \\
p(R) 
\end{array} \right]^{\otimes n}
\begin{array}{c}
n\rightarrow \! \infty 
\end{array}\!\!$ $\blankthreearray$
& $n(1\!\!-\!\!H)kT\ln 2$ & $n(1\!\!-\!\!H)kT\ln 2$\\
\hline 
$\left[\begin{array}{c}
0.7 \\
0.3 
\end{array} \right]^{\otimes 1000}\!\!\!\! T\!\!=\!\!T_{room} $ $\blankthreearray$& 1.0 eV  & 3.5 eV \\

\hline 
$\frac{1}{2}\left[\begin{array}{c}
1 \\
0 
\end{array} \right]^{\otimes n}\!\!\!\!+\! \frac{1}{2}\left[\begin{array}{c}
1/2 \\
1/2 
\end{array} \right]^{\otimes n} $ $\blankthreearray$& $\approx 0$ & $\approx nkT\ln2$\\

\hline 
$\frac{1}{2}\left[\begin{array}{c}
1 \\
0 
\end{array} \right]^{\otimes n}\!\!+ \frac{1}{2}\left[\begin{array}{c}
0 \\
1 
\end{array} \right]^{\otimes n} $ $\blankthreearray$& $\approx nkT\ln2$ & $\approx nkT\ln2$\\

\hline 
\end{tabular}
\caption{Examples of distributions and their work values. We set the work value of a box $c=kT\ln2$. The notation $\left[(.)\right]^{\otimes n}$ means that the distribution is combined with itself independently $n$ times. The first two examples are discussed more in Figure~\ref {fig:convergenceisslow}. The third example shows that the statements are both needed as they do not in general approximate one another. The fourth distribution is an example where access to the information compressing unitary leads to an almost maximal amount of work being extractable.}

\end{table}


It is moreover interesting that the standard thermodynamical heat engine work extraction scheme is a strategy amongst those considered here, corresponding to the unitary being the identity and all bits being selected. In general this is a suboptimal strategy (it could for example not extract any work given example four), although in the thermodynamical limit (of the first example above) it is, interestingly, optimal. This is because the standard thermodynamical work from $P(L)n$ ($P(R)n$) particles on the left (right) of the divider can be shown to be given by $W=n(1-H_S)kT\ln2$.  This quantity is identical to the extractable work if one also allows the information compressing unitary ---see the first example in Table~1. 
\begin{figure}
\centering \includegraphics[width=0.8\linewidth]{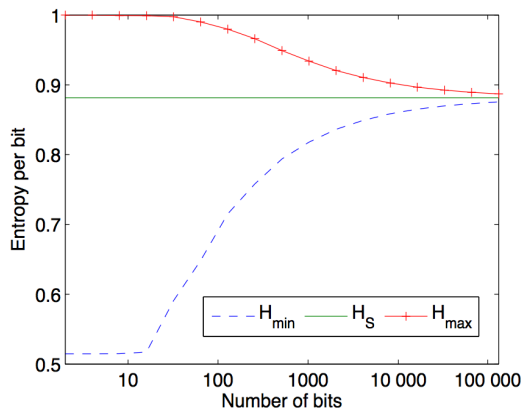}
\caption{Three entropies, $H_{\min}$, $H_S$ and $H_{\max}$  are evaluated for $n$ uncorrelated bits with $p(L)=0.7$ (a choice motivated by the experiment \cite{SerreliLKL07}). One sees that in the $n\to \infty$ limit the entropies coincide. This is because all bits are in this limit (after an appropriate unitary) uniformly random or fully known. Accordingly the first example in Table~1 has the same amount of `min' and `max' work. The figure also shows that $n$ needs to be large in general for the entropies to coincide, which is why the second example in Table~1 shows that the type I and II work extractable are significantly different for $n=1000$. For more details on this figure, see~\cite{Figurenotes}.}
\label{fig:convergenceisslow}
\end{figure}

Whilst this type of work extraction is experimentally more challenging than standard thermodynamical work extraction, it is nevertheless significantly easier than full quantum computation. One will in general not require a universal set of unitaries nor a high accuracy in order to demonstrate non-trivial work extraction (or `resetting' which is the inverse of that~\cite{Feynman}) via information compression. It seems likely that at least some of the multitude of methods being developed for performing quantum gates will be suitable for this purpose, and that they will be sufficiently good for this application significantly earlier than they can be used for quantum computing~\cite{QC}. Similar experiments have already been performed in the context of NMR algorithmic cooling~\cite{FernandezLMR04, BaughMRNL04}.

We finally stress that two agents with differing knowledge about the same system and/or differing risk tolerance can extract different amounts of work; the extractable work must be seen as {\em subjective} in that sense.   

The results of this work will be presented in detail in forthcoming publications.

{\bf {\em Conclusion---}}  We have quantified the relation between work and information, employing the smooth-entropy approach. We suggested work extraction is a guessing game where the amount of work an agent can extract depends both on its knowledge and its risk tolerance. We nevertheless recovered a simple relation between work and information by noting that all risk tolerances are in between 0 and arbitrary risk, and deriving the work value of information for those two cases. The results point the way to achieving for statistical mechanics what smooth entropies accomplished for information theory more generally. A natural next step is to apply our approach to related work/information scenarios, including NMR algorithmic cooling.

We gratefully acknowledge discussions with J. Aaberg, R. Colbeck, M. Gu, A. Imamoglu, D. Kaszlikowski, S. Lloyd, L. del Rio and S. Winkler. OD and RR were funded by ETH Z\"urich. OD is also grateful to the Singapore Centre for Quantum Technologies for hosting him for a month during this work. ER was funded by Singapore Centre for Quantum Technologies. VV is grateful to EPSRC, the Royal Society, the Wolfson Foundation, the National Research Foundation (Singapore), and the Ministry of Education (Singapore) for financial support.


\begin{thebibliography}{99}
\bibitem{Figurenotes}
In Figure \ref{fig:convergenceisslow} the smoothing is implicit, and $\epsilon=10^{-5}$.  The kink location in the $H_{\min}$ curve is because the smoothing has little impact when $p_{\max}=0.7^n \gg \epsilon$.
\end{thebibliography}
\end{document}